\documentclass[runningheads]{paper648}

\usepackage{multirow}
\usepackage{amsmath,graphicx,booktabs,amssymb,stmaryrd}
\usepackage{wrapfig}
\usepackage{subcaption}
\usepackage{caption}
\usepackage{color}
\usepackage{bbm}
\usepackage{booktabs}
\usepackage{bbding}
\usepackage[pagebackref=true,breaklinks=false,colorlinks,bookmarks=false]{hyperref}
\usepackage{algorithm}
\usepackage{algpseudocode}

\usepackage[dvipsnames]{xcolor}

\begin{document}

\title{Pseudo Bias-Balanced Learning for Debiased Chest X-ray Classification}

\author{Luyang Luo\inst{1}\Envelope, 
Dunyuan Xu\inst{1},
Hao Chen\inst{2}, \\
Tien-Tsin Wong\inst{1}, 
\and Pheng-Ann Heng\inst{1}}
% index{Luo, Luyang}
% index{Xu, Dunyuan}
% index{Chen, Hao}
% index{Wong, Tien-Tsin}
% index{Heng, Pheng-Ann}

\institute{$^1$Department of Computer Science and Engineering,\\
The Chinese University of Hong Kong, Hong Kong, China\\
\email{lyluo@cse.cuhk.edu.hk}\\
$^2$Department of Computer Science and Engineering,\\
The Hong Kong University of Science and Technology, Hong Kong, China}

\authorrunning{Luyang Luo, et al.}

\titlerunning{Pseudo Bias-Balanced Learning for Debiased Chest X-ray Classification}

\maketitle              

%\footnotetext[1]{}

\begin{abstract}
{
Deep learning models were frequently reported to learn from shortcuts like dataset biases.
As deep learning is playing an increasingly important role in the modern healthcare system, it is of great need to combat shortcut learning in medical data as well as develop unbiased and trustworthy models.
In this paper, we study the problem of developing debiased chest X-ray diagnosis models from the biased training data without knowing exactly the bias labels.
We start with the observations that the imbalance of bias distribution is one of the key reasons causing shortcut learning, and the dataset biases are preferred by the model if they were easier to be learned than the intended features.
Based on these observations, we proposed a novel algorithm, pseudo bias-balanced learning, which first captures and predicts per-sample bias labels via generalized cross entropy loss and then trains a debiased model using pseudo bias labels and bias-balanced softmax function.
%To our best knowledge, we are pioneered in tackling dataset biases in medical images without explicit labeling on the bias attributes.
We constructed several chest X-ray datasets with various dataset bias situations and demonstrated with extensive experiments that our proposed method achieved consistent improvements over other state-of-the-art approaches.\footnote[1]{Code available at \url{https://github.com/LLYXC/PBBL}.}
}

\end{abstract}

\section{Introduction}

To date, deep learning (DL) has achieved comparable or even superior performance to experts on many medical image analysis tasks \cite{rajpurkar2022ai}.
Robust and trustworthy DL models are hence of greater need than ever to unleash their huge potential in solving real-world healthcare problems.
However, a common trust failure of DL was frequently found where the models reach a high accuracy without learning from the intended features.
For example, using backgrounds to distinguish foreground objects \cite{ribeiro2016should}, using the gender to classify hair colors \cite{sagawa2020distributionally}, or worse yet, using patients' position to determine COVID-19 pneumonia from chest X-rays \cite{degrave2021ai}.
Such a phenomenon is called \emph{shortcut learning} \cite{geirhos2020shortcut}, where the DL models choose unintended features, or \emph{dataset bias}, for making decisions.

More or less, biases could be generated during the creation of the datasets \cite{torralba2011unbiased}.
As the dataset biases frequently co-occurred with the primary targets, the model might take shortcuts by learning from such spurious correlation to minimize the empirical risk over the training data.
As a result, dramatic performance drops could be observed when applying the models onto other data which do not obtain the same covariate shift \cite{luo2021rethinking}.
In the field of medical image analysis, shortcut learning has also been frequently reported, such as using hospital tokens to recognize pneumonia cases \cite{zech2018variable}; learning confounding patient and healthcare variables to identify fracture cases; relying on chest drains to classify pneumothorax case \cite{oakden2020hidden}; or leveraging shortcuts to determine COVID-19 patients \cite{degrave2021ai}.
These findings reveal that shortcut learning makes deep models less explainable and less trustworthy to doctors as well as patients, and addressing shortcut learning is a far-reaching topic for modern medical image analysis.

To combat shortcut learning and develop debiased models, a branch of previous works use data re-weighting to learn from less biased data.
For instance, REPAIR \cite{li2019repair} proposed to solve a minimax problem between the classifier parameters and dataset re-sampling weights.
Group distributional robust optimization \cite{sagawa2020distributionally} prioritized worst group learning, which was also mianly implemented by data re-weighting.
Yoon et al. \cite{yoon2019generalizable} proposed to address dataset bias with a weighted loss and a dynamic data sampler.
Another direction of works emphasizes learning invariance across different environments, such as invariant risk minimization \cite{arjovsky2019invariant}, contrastive learning \cite{tartaglione2021end}, and mutual information minimization \cite{zhu2021learning}.
However, these methods all required dataset biases to be explicitly annotated, which might be infeasible for realistic situations, especially for medical images.
Recently, some approaches have made efforts to relax the dependency on explicit bias labels.
Nam et al. \cite{nam2020learning} proposed to learn a debiased model by mining the high-loss samples with a highly-biased model.
Lee et al. \cite{lee2021learning} further incorporated feature swapping between the biased and debiased models to augment the training samples.
Yet, very few methods attempted to efficiently address shortcut learning in medical data without explicitly labeling the biases.

In this paper, we are pioneered in tackling the challenging problem of developing debiased medical image analysis models without explicit labels on the bias attributes.
We first observed that the imbalance of bias distribution is one of the key causes to shortcut learning, and dataset biases would be preferred when they were easier to be learned than the intended features.
We thereby proposed a novel algorithm, namely pseudo bias-balanced learning (PBBL).
PBBL first develops a highly-biased model by emphasizing learning from the easier features.
The biased model is then used to generate pseudo bias labels that are later utilized to train a debiased model with a bias-balanced softmax function.
We constructed several chest X-ray datasets with various bias situations to evaluate the efficacy of the debiased model.
We demonstrated that our method was effective and robust under all scenarios and achieved consistent improvements over other state-of-the-art approaches.

\section{Methodology}

\subsection{Problem Statement and Study Materials}
Let $X$ be the set of input data, $Y$ the set of target attributes that we want the model to learn, and $B$ the set of bias attributes that are irrelevant to the targets.
Our goal is to learn a function $f:X\rightarrow Y$ that would not be affected by the dataset bias.
We here built the following chest X-ray datasets for our study.

\textbf{Source-biased Pneumonia (SbP): }
For the training set, we first randomly sampled 5,000 pneumonia cases from MIMIC-CXR \cite{johnson2019mimic} and 5,000 healthy cases (no findings) from NIH \cite{wang2017chestx}.
We then sampled $5,000\times r\%$ pneumonia cases from NIH and the same amount of healthy cases from MIMIC-CXR.
Here, the \texttt{data source} became the dataset bias, and \texttt{health condition} was the target to be learned.
We varied $r$ to be 1, 5, and 10, which led to biased sample ratios of 99\%, 95\%, and 90\%, respectively.
We created the validation and the testing sets by equally sampling 200 and 400 images from each group (w/ or w/o pneumonia; from NIH or MIMIC-CXR), respectively.
Moreover, as overcoming dataset bias could lead to better external validation performance \cite{geirhos2020shortcut}, we included 400 pneumonia cases and 400 healthy cases from Padchest \cite{bustos2020padchest} to evaluate the generalization capability of the proposed method.
Note that we converted all images to JPEG format to prevent the data format from being another dataset bias.

\textbf{Gender-biased Pneumothorax (GbP): }
Previous study \cite{larrazabal2020gender} pointed out that gender imbalance in medical datasets could lead to a biased and unfair classifier.
Based on this finding, we constructed two training sets from the NIH dataset \cite{wang2017chestx}: 
1) \textbf{GbP-Tr1}: 800 male samples with pneumothorax, 100 male samples with no findings, 800 female samples with no findings, and 100 female samples with pneumothorax; 
2)\textbf{GbP-Tr2}: 800 female samples with pneumothorax, 100 female samples with no findings, 800 male samples with no findings, and 100 male samples with pneumothorax.
For validation and testing sets, we equally collected 150 and 250 samples from each group (w/ or w/o pneumothorax; male or female), respectively.
Here, \texttt{gender} became a dataset bias and \texttt{health condition} was the target that the model was aimed to learn.

Following previous studies \cite{nam2020learning,lee2021learning}, we call a sample bias-aligned if its target and bias attributes are highly-correlated in the training set (e.g., (\texttt{pneumonia, MIMIC-CXR}) or (\texttt{healthy, NIH}) in the SbP dataset).
On the contrary, a sample is said to be bias-conflicting if the target and bias attributes are dissimilar to the previous situation (e.g., (\texttt{pneumonia, NIH}) or (\texttt{healthy, MIMIC-CXR})).

\begin{figure}[t]
\centering
\begin{subfigure}{.34\textwidth}
  \centering
  \includegraphics[width=0.9\linewidth]{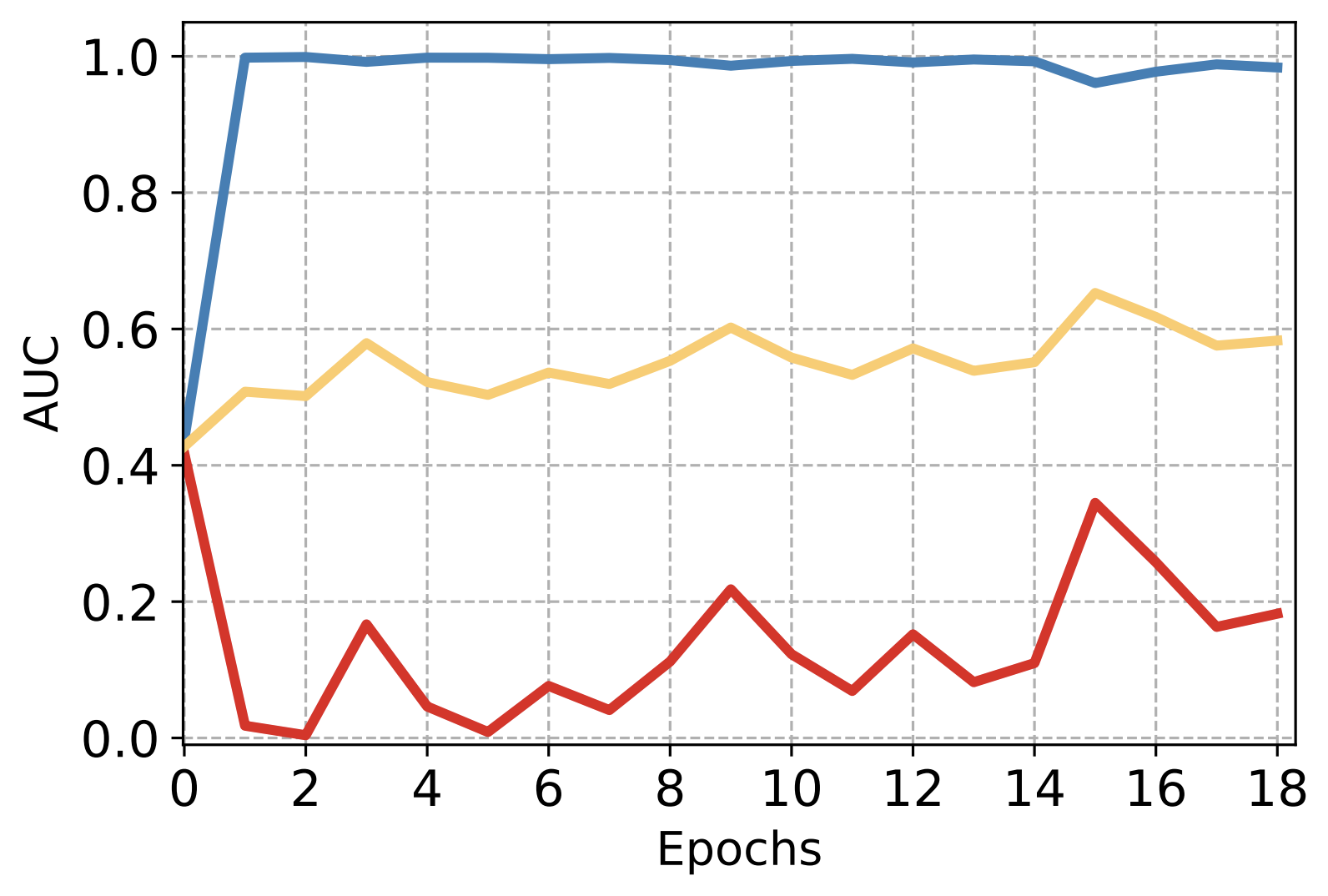}
  \caption{}
  \label{bias_balance:sfig1}
\end{subfigure}
\begin{subfigure}{.34\textwidth}
  \centering
  \includegraphics[width=0.9\linewidth]{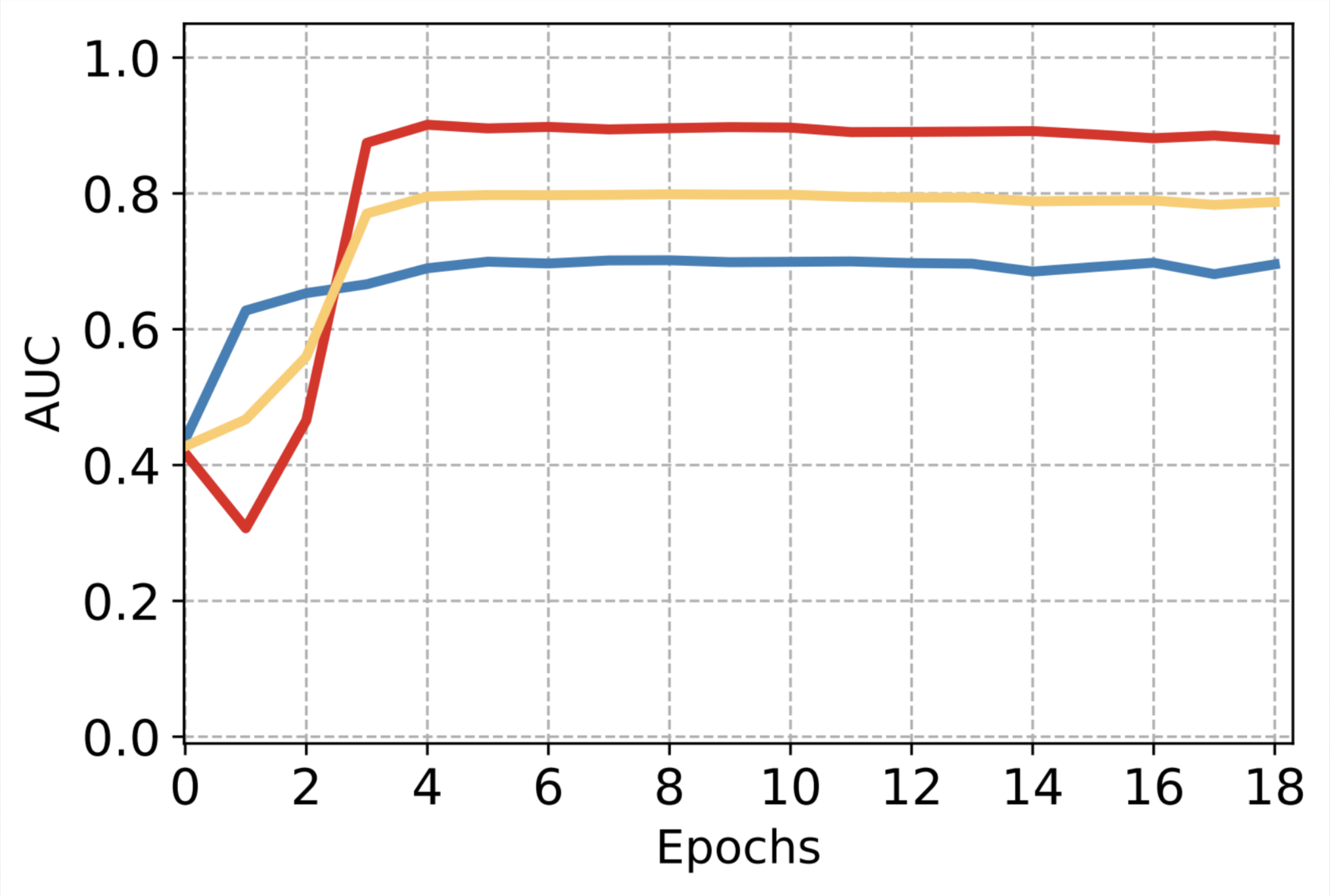}
  \caption{}
  \label{bias_balance:sfig2}
\end{subfigure}
\begin{subfigure}{.26\textwidth}
  \centering
  \includegraphics[width=0.83\linewidth]{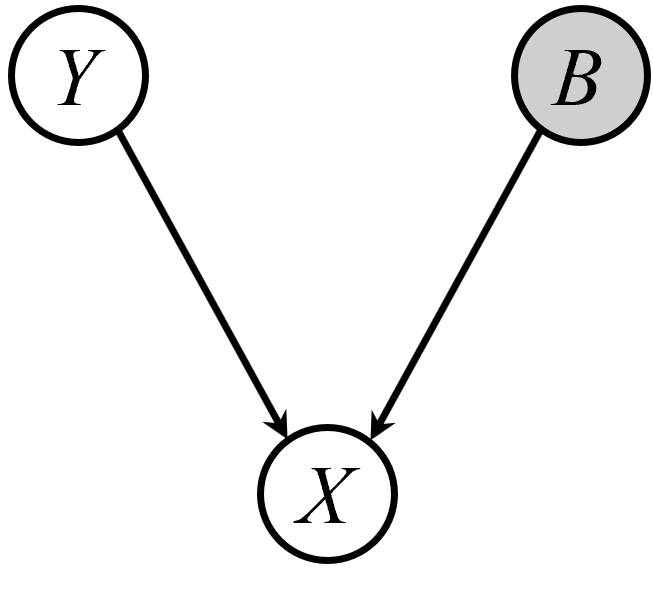}
  \caption{}
  \label{bias_balance:sfig3}
\end{subfigure}
\caption{We show (a) the testing results in AUC curves of a model trained on Source-biased Pneumonia dataset; (b) the testing results in AUC curves of a model trained with bias-balanced pneumonia dataset. We further show our causal assumption of data generation process in (c). \textcolor{Cerulean}{Blue curves:} results on bias-aligned samples; \textcolor{BrickRed}{Red curves:} results on bias-conflicting samples; \textcolor{Goldenrod}{Yellow curves:} averaged results of bias-aligned AUC and bias-conflicting AUC.}
\label{bias_balance:fig1}
\end{figure}

\subsection{Bias-balanced Softmax}

Our first observation is that \emph{bias-imbalanced training data leads to a biased classifier}.
Based on the SbP dataset, we trained two different settings: i) SbP with $r=10$; ii) Bias balancing by equally sampling 500 cases from each group.
The results are shown in Fig. \ref{bias_balance:sfig1} and Fig. \ref{bias_balance:sfig2}, respectively.
Clearly, when the dataset is bias-imbalanced, learning bias-aligned samples were favored.
On the contrary, balancing the biases mitigates shortcut learning even with less training data.

For a better interpretation, we adopt the causal assumption \cite{mitrovic2020representation} that the data $X$ is generated from both the target attributes $Y$ and the bias attributes $B$, which are independent to each other, as shown in Fig. \ref{bias_balance:sfig3}.
The conditional probability $p(y=j|x)$ hence can be formalized as follows:
\begin{equation}
    p(y=j|x,b) = \frac{p(x|y=j,b)p(y=j|b)}{p(x|b)},
\end{equation}
where $p(y=j|b)$ raises a distributional discrepancy between the biased training data and the ideal bias-balanced data (e.g., the testing data).
Moreover, according to our experimental analysis before, the imbalance also made the model favor learning from bias-aligned samples, which finally resulted in a biased classifier.
To tackle the bias-imbalance situation, let $k$ be the number of classes and $n_{j,b}$ the number of training data of target class $j$ with bias class $b$, we could derive a bias-balanced softmax \cite{hong2021unbiased,ren2020balanced} as follows:

\begin{theorem}
(Bias-balanced softmax \cite{hong2021unbiased})
Assume $\phi_{j} = p(y=j|x,b)=\frac{p(x|y=j,b)}{p(x|b)}\cdot\frac{1}{k}$ to be the desired conditional probability of the bias-balanced dataset, and $\hat{\phi}_{j}=\frac{p(x|y=j,b)}{\hat{p}(x|b)}\cdot\frac{n_{j,b}}{\sum_{i=1}^{k}n_{i,b}}$ to be the conditional probability of the biased dataset. If $\phi$ can be expressed by the standard Softmax function of the logits $\eta$ generated by the model, i.e., $\phi_{j}=\frac{exp(\eta_{j})}{\sum_{i=1}^{k}exp(\eta_{i})}$, then $\hat{\phi}$ can be expressed as
\begin{equation}
    \hat{\phi}_{j} = \frac{p(y=j|b)\cdot {\rm exp}(\eta_{j})}{\sum_{i=1}^{k}p(y=i|b)\cdot {\rm exp}({\eta_{i}})}.
\end{equation}
\label{Theorem_bias_balanced_softmax}
\end{theorem}

Theorem \ref{Theorem_bias_balanced_softmax} (proof provided in the supplementary) shows that bias-balanced softmax could well solve the distributional discrepancy between the bias-imbalanced training set and the bias-balanced testing set.
Denoting $M$ the number of training data, we obtain the bias-balanced loss for training a debiased model:
\begin{equation}
    \mathcal{L}_{\rm BS}(f(x), y, b) = -\frac{1}{M}\sum_{i=1}^{M}log\Bigg(\frac{p(y=j|b)\cdot {\rm exp}(\eta_{j})}{\sum_{i=1}^{k}p(y=i|b)\cdot {\rm exp}({\eta_{i}})}\Bigg)
\end{equation}

However, this loss requires estimation of the bias distribution on the training set, while comprehensively labeling all kinds of attributes would be unpractical, especially for medical data.
In the next section, we elaborate on how to obtain the estimation of the bias distribution without knowing the bias labels.

\subsection{Bias Capturing with Generalized Cross Entropy Loss}

\begin{figure}[h]
\centering
\begin{subfigure}{.32\textwidth}
  \centering
  \includegraphics[width=0.9\linewidth]{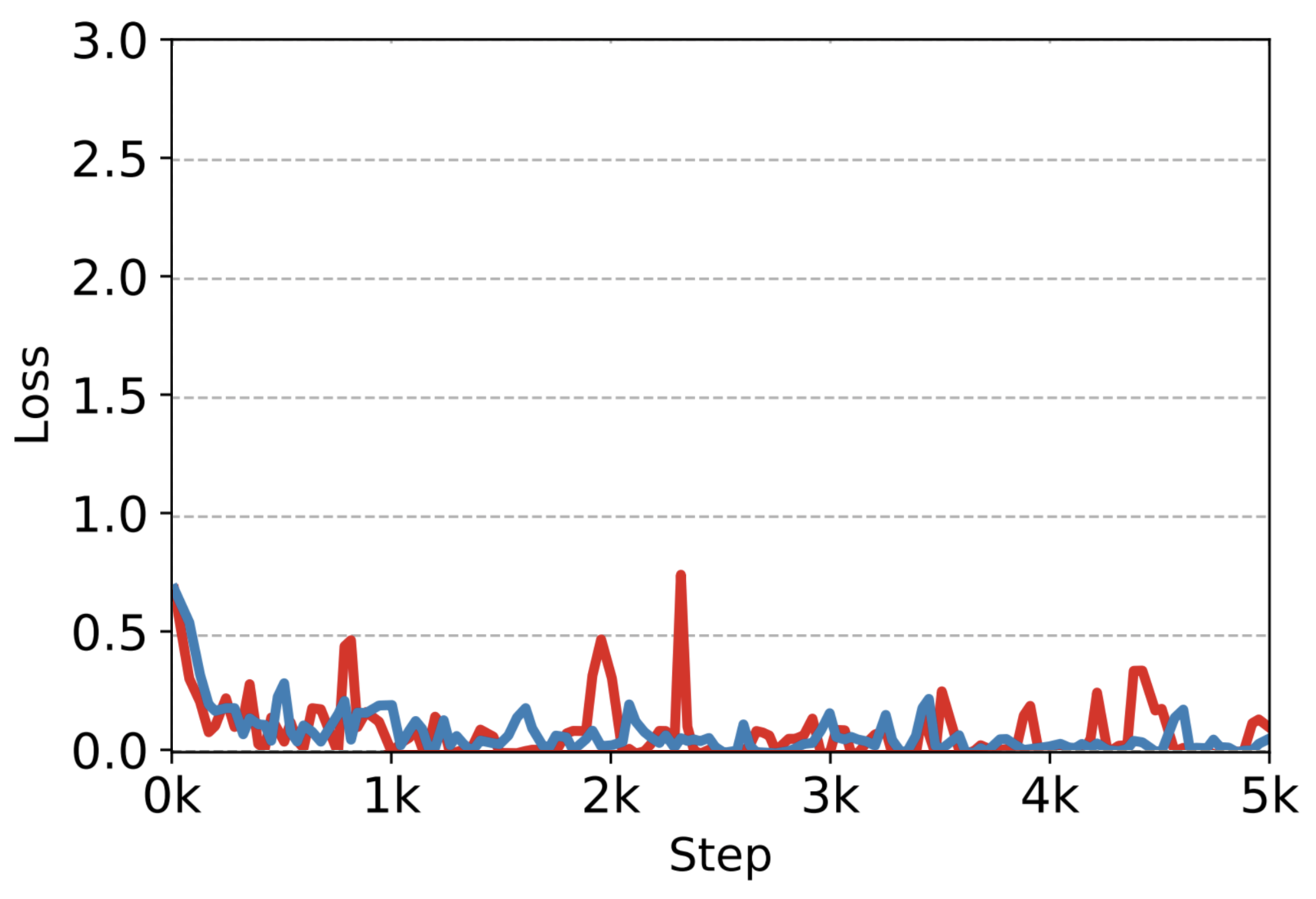}
  \caption{}
  \label{loss_curve:sfig1}
\end{subfigure}
\begin{subfigure}{.32\textwidth}
  \centering
  \includegraphics[width=0.9\linewidth]{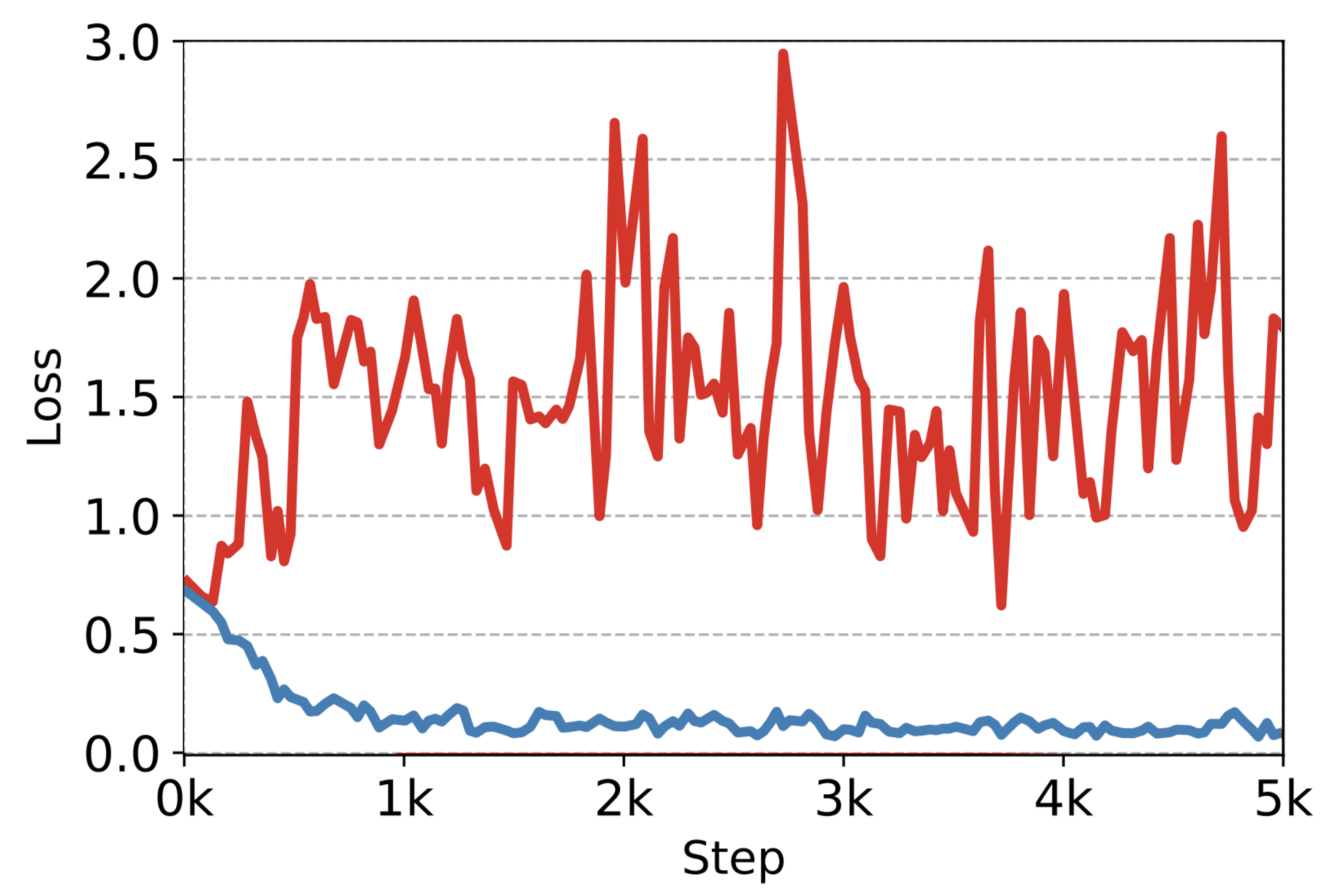}
  \caption{}
  \label{loss_curve:sfig2}
\end{subfigure}
\begin{subfigure}{.32\textwidth}
  \centering
  \includegraphics[width=0.9\linewidth]{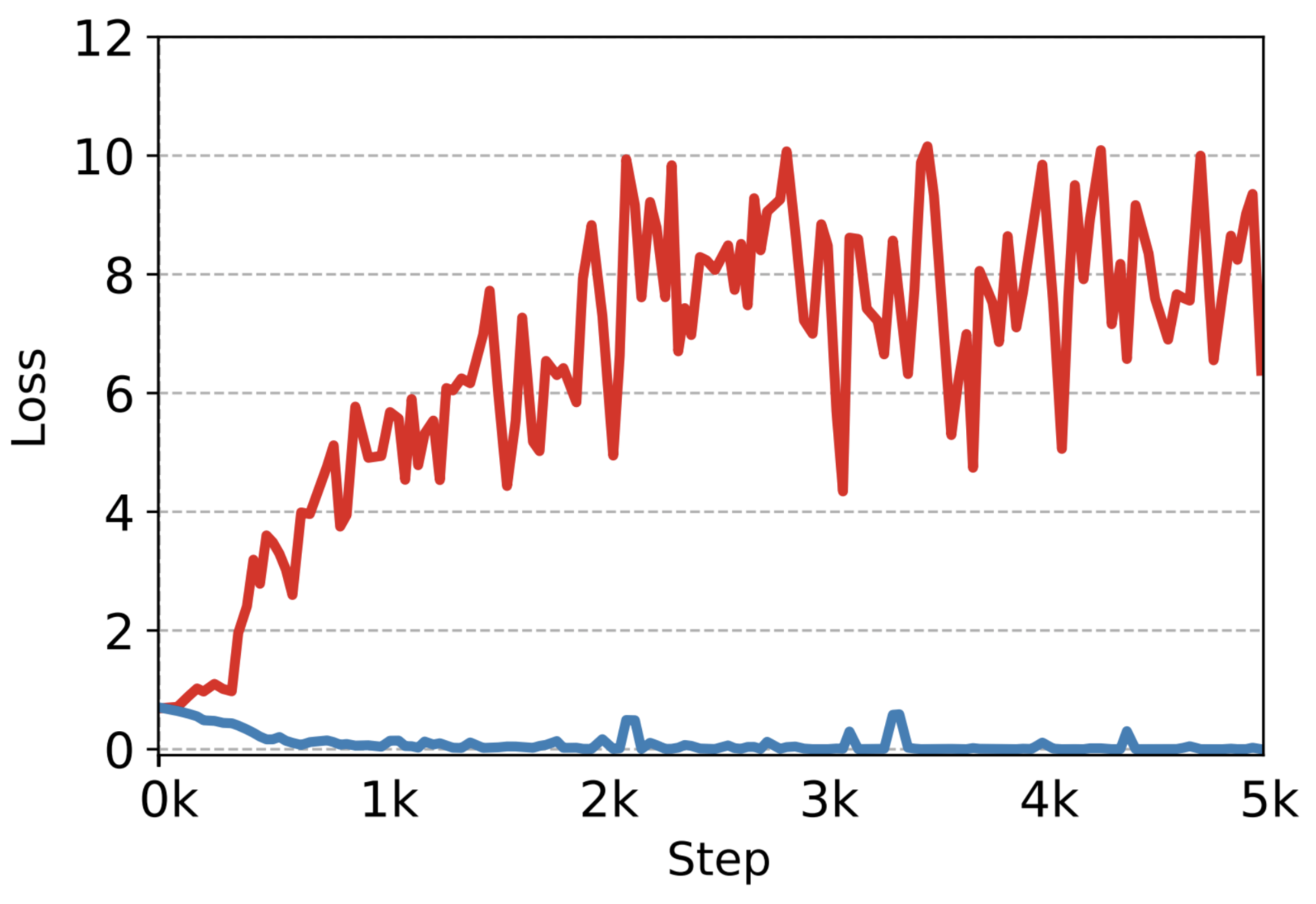}
  \caption{}
  \label{loss_curve:sfig3}
\end{subfigure}
\caption{
Based on the SbP dataset, we show the learning curve of the vanilla model by setting the \{\texttt{target, bias}\} pair to be (a) \{\texttt{data source, health condition}\} and (b) \{\texttt{health condition, data source}\}. We also show in (c) the learning curve of a highly-biased model trained with GCE loss with the \{\texttt{target, bias}\} pair being \{\texttt{health condition, data source}\}. \textcolor{Cerulean}{Blue curves}: loss of bias-aligned samples; \textcolor{BrickRed}{Red curves}: loss of bias-conflicting samples.}
\label{loss_curve:fig1}
\end{figure}

Inspired by \cite{nam2020learning}, we conducted two experiments based on the Source-biased Pneumonia dataset with $r=10$, where we set the models to classify data source (Fig. \ref{loss_curve:sfig1}) or health condition (Fig. \ref{loss_curve:sfig2}), respectively.
Apparently, the model has almost no signs of fitting on the bias attribute (health condition) when it's required to distinguish data source.
%even when each data source is highly correlated with different health conditions
On the other hand, the model quickly learns the biases (data source) when set to classify pneumonia from healthy cases.
From these findings, one could conclude that \emph{dataset biases would be preferred when they were easier to be learned than the intended features}.

Based on this observation, we could develop a model to capture the dataset bias by making it quickly fit on the easier features from the training data.
Therefore, we adopt the generalized cross entropy (GCE) loss \cite{zhang2018generalized}, which was originally proposed to address noisy labels by fitting on the easier clean data and slowly memorizing the hard noisy samples.
Inheriting this idea, the GCE loss could also quickly capture easy and biased samples than the categorical cross entropy (CE) loss.
Giving $f(x;\theta)$ the softmax output of the model, denoting $f_{y=j}(x;\theta)$ the probability of $x$ being classified to class $y=j$ and $\theta$ the parameters of model $f$, the GCE loss is formulated as follows:
\begin{equation}
    \mathcal{L}_{\rm GCE}(f(x;\theta), y=j) = \frac{1-f_{y=j}(x;\theta)^{q}}{q},
\end{equation}
where $q$ is a hyper-parameter.
The gradient of GCE is $\frac{\partial \mathcal{L}_{\rm GCE}(f(x;\theta), y=j)}{\partial \theta} = f_{y=j}(x;\theta)^{q}\frac{\partial \mathcal{L}_{\rm CE}(f(x;\theta), y=j)}{\partial\theta}$ (proof provided in the supplementary), which explicitly assigns weights on the CE loss based on the agreement between model's predictions and the labels.
As shown in Fig. \ref{loss_curve:sfig3}, GCE loss fits the bias-aligned samples quickly while yields much higher loss on the bias-conflicting samples.

\subsection{Bias-balanced Learning with Pseudo Bias}

With the afore discussed observations and analysis, we propose a debiasing algorithm, namely Pseudo Bias Balanced Learning.
We first train a biased model $f_B(x;\theta_{B})$ with the GCE loss and calculate the corresponding receiver operating characteristics (ROC) over the training set.
Based on the ROC curve, we compute the sensitivity $u(\tau)$ and specificity $v(\tau)$ under each threshold $\tau$ and then assign pseudo bias labels to each sample with the following:
\begin{equation}
    \tilde{b}(f_B(x; \theta_{B})) =\begin{cases}
                        1,   & \text{if } f_B(x; \theta_{B})\geq {\rm argmax}_{\tau}(u(\tau)+v(\tau));\\
                        0,   & \text{otherwise}.
                    \end{cases}
\label{bias_assign}
\end{equation}

\begin{algorithm}[h!]
    \caption{Pseudo Bias Balanced Learning}\label{PBBL}
    \hspace*{\algorithmicindent} \textbf{Input:} $\theta_{B}$, $\theta_{D}$, image $x$, target label $y$, numbers of iterations $T_{B}$, $T_{D}$, $N$.\\
    \hspace*{\algorithmicindent} \textbf{Output:} Debiased model $f_{D}(x; \theta_{D})$.
    \begin{algorithmic}[1]
    \State Initialize $\tilde{b} = y$.
    \For{n=1, $\cdots$, $N$}
        \State Initialize network $f_B(x; \theta_{B})$.
        \For{t=1, $\cdots$, $T_{B}$} 
            \State{Update $f_{B}(x; \theta_{B})$ with $\mathcal{L}_{\rm GCE}(f_{B}(x;\theta_{B}), \tilde{b})$}
        \EndFor
        \State Calculate $u$, $v$, and $\tau$ over training set.
        \State Update pseudo bias labels $\tilde{b}$ with Eq. \ref{bias_assign}.
    \EndFor
    \State Initialize network $f_D(x;\theta_{D})$.
    \For{t=1, $\cdots$, $T_{D}$} 
        \State{Update $f_{D}(x; \theta_{D})$ with $\mathcal{L}_{\rm BS}(f_{D}(x;\theta_{D}), y, \tilde{b})$}
    \EndFor
    
    \end{algorithmic}
\end{algorithm}

Moreover, as the biased model could also memorize the correct prediction for the hard bias-conflicting cases \cite{zhang2017understanding}, we propose to capture and enhance the bias via iterative model training.
Finally, we train our debiased model $f_D(x;\theta_{D})$ based on the pseudo bias labels and the bias-balance softmax function, with different weights from $\theta_{B}$.
The holistic approach is summarized in Algorithm \ref{PBBL}.

\section{Experiments}
 
\textbf{Evaluation metrics} are the area under the ROC curve (AUC) with four criteria: i) AUC on bias-aligned samples; ii) AUC on bias-conflicting samples; iii) Average of bias-aligned AUC and bias-conflicting AUC, which we call balanced-AUC; iv) AUC on all samples.
The difference between the first two metrics could reflect whether the model is biased, while the latter two metrics provide unbiased evaluations on the testing data.

\textbf{Compared methods} included four other approaches: i) Vanilla model, which did not use any debiasing strategy and could be broadly regarded as a lower bound. ii) Group Distribution Robust Optimization (G-DRO) \cite{sagawa2020distributionally}, which used the bias ground truth and could be regarded as the upper bound. G-DRO divides training data into different groups according to their targets and bias labels. It then optimized the model with priority on the worst-performing group and finally achieved robustness on every single group. As in practical scenarios, the labels for the dataset biases may not be known, we also implemented iii) Learning from Failure (LfF) \cite{nam2020learning}, which developed a debiased model by weighted losses from a biased model; and iv) Disentangled Feature Augmentation (DFA) \cite{lee2021learning}, which was based on LfF and further adds feature swapping and augmentation between the debiased and biased models.

\textbf{Model training protocol} is as follows: We used the same backbone, DenseNet-121 \cite{huang2017densely} with pre-trained weights from \cite{cohen2021torchxrayvision}, for every method.
Particularly, we fixed the weights of DenseNet, replaced the final output layer with three linear layers, and used the rectified linear units as the intermediate activation function.
We ran each model with three different random seeds, and reported the test results corresponding to the best validation AUC.
Each model is optimized with Adam \cite{kingma2014adam} for around 1,000 steps with batch size of 256 and learning rate of 1e-4. 
$N$ in Algorithm \ref{PBBL} is empirically set to 1 for SbP dataset and 2 for GbP dataset, respectively. $q$ in GCE loss is set to 0.7 as recommended in \cite{zhang2018generalized}.

\begin{table}[h]
\centering
\caption{AUC results on SbP dataset. Best results without ground truth bias labels are emphasized in \textbf{bold}. $\dagger$ means the method uses ground truth bias labels.}
\scalebox{0.8}{\begin{tabular}{c|c|cccc|c}
\hline
\textbf{Bias Ratio}                                 &  \textbf{ Method}    & \textbf{Aligned} & \textbf{Conflicting} & \textbf{Balanced} & \textbf{Overall}                      & \textbf{External}  \\ \hline
{\multirow{5}{*}{90\%}}                      & {G-DRO$^{\dagger}$ \cite{sagawa2020distributionally}}   & 70.02$_{\pm2.20}$ & {89.80$_{\pm0.87}$}& 79.94$_{\pm0.68}$& 80.23$_{\pm0.37}$ & 90.06$_{\pm0.32}$ \\

{} & {Vanilla} & \textbf{96.51$_{\pm0.26}$} & {31.21$_{\pm3.04}$} & 63.86$_{\pm1.39}$& 69.84$_{\pm1.32}$ & 71.57$_{\pm0.90}$ \\
{} & {LfF \cite{nam2020learning}}   & 68.57$_{\pm2.16}$ & \textbf{87.46$_{\pm2.17}$}& 78.02$_{\pm0.18}$& 78.26$_{\pm0.18}$ & 87.71$_{\pm2.66}$ \\
{}   & {DFA \cite{lee2021learning}}  & 74.63$_{\pm4.61}$ & {83.30$_{\pm3.96}$}  & 78.96$_{\pm0.33}$ & 78.76$_{\pm0.15}$ & 74.58$_{\pm7.56}$ \\
{}    & {Ours}   & 76.82$_{\pm2.80}$  & {85.75$_{\pm0.32}$}& \textbf{80.49}$_{\pm0.20}$& \textbf{78.78$_{\pm3.02}$}  & \textbf{89.96$_{\pm0.69}$} \\ \hline
{\multirow{5}{*}{95\%}}                      & {G-DRO$^{\dagger}$ \cite{sagawa2020distributionally}}  & 68.65$_{\pm1.21}$ & {89.86$_{\pm0.67}$}  & 79.26$_{\pm0.47}$& 79.8$_{\pm0.36}$& 90.16$_{\pm0.73}$ \\

{}  & {Vanilla} & \textbf{97.91$_{\pm0.75}$}  & {20.45$_{\pm5.96}$} & 59.18$_{\pm2.61}$& 67.11$_{\pm1.85}$& 68.61$_{\pm3.50}$ \\
{}  & {LfF \cite{nam2020learning}}     & 69.56$_{\pm2.01}$ & \textbf{86.43$_{\pm1.67}$}& 77.99$_{\pm0.18}$& \textbf{78.28$_{\pm0.22}$} & \textbf{88.56$_{\pm3.37}$} \\

{} & {DFA \cite{lee2021learning}}     & 69.04$_{\pm4.21}$ & 84.94$_{\pm2.56}$ & 76.99$_{\pm0.85}$& 77.26$_{\pm0.49}$ & 76.37$_{\pm3.26}$ \\
{}     & {Ours}   & 71.72$_{\pm6.65}$ & {84.68$_{\pm3.49}$}  & \textbf{78.20$_{\pm0.20}$} & 78.04$_{\pm3.46}$ & 82.65$_{\pm0.40}$ \\ \hline
{\multirow{5}{*}{99\%}}                      & {G-DRO$^{\dagger}$ \cite{sagawa2020distributionally}}  & 74.30$_{\pm2.28}$ & {85.18$_{\pm1.26}$}  & 79.74$_{\pm0.55}$& 79.71$_{\pm0.40}$ & 89.87$_{\pm0.64}$ \\
{}  & {Vanilla} & \textbf{99.03$_{\pm0.95}$} & {4.93$_{\pm3.68}$} & 51.98$_{\pm1.60}$& 59.21$_{\pm3.76}$ & 60.79$_{\pm0.98}$ \\
{} & {LfF \cite{nam2020learning}}   & 77.50$_{\pm11.08}$ & {64.38$_{\pm8.75}$} & 70.94$_{\pm1.30}$& 71.86$_{\pm1.72}$ & 73.90$_{\pm4.42}$ \\
{}  & {DFA \cite{lee2021learning}}   & 69.33$_{\pm1.74}$ & 75.48$_{\pm2.61}$ & 72.40$_{\pm0.48}$& 72.49$_{\pm0.45}$& 61.67$_{\pm6.86}$ \\

{} & {Ours}    & 72.40$_{\pm0.71}$ & \textbf{77.61$_{\pm0.45}$} & \textbf{75.00$_{\pm0.18}$} & \textbf{74.70$_{\pm0.14}$} & \textbf{78.87$_{\pm0.44}$} \\ \hline
\end{tabular}}
\label{SbP_comparison}
\end{table}

\textbf{Quantitative results on Source-biased Pneumonia dataset} are reported in Table \ref{SbP_comparison}.
With the increasing of bias ratio, the vanilla model became more and more biased and severe decreases in balanced-AUC and overall-AUC was observed.
All other methods also showed decreases on the two metrics, while G-DRO shows quite robust performance under all situations.
Meanwhile, our method achieved consistent improvement over the compared approaches under most of the situations, demonstrating its effectiveness in debiasing.
Interestingly, the change of external testing performance appeared to be in line with the change of the balanced-AUC and overall AUC, which further revealed that overcoming shortcut learning improves the model's generalization capability.
These findings demonstrated our method's effectiveness in solving shortcut learning, with potential in robustness and trustworthiness for real-world clinic usage.

\begin{table}[t]
\centering
\caption{AUC results on GbP dataset. Best results without ground truth bias labels are emphasized in \textbf{bold}. $\dagger$ means the method uses ground truth bias labels.}
\scalebox{0.8}{\begin{tabular}{c|c|cccc}
\hline
\textbf{Training}  &  \textbf{ Method}  & \textbf{Aligned} & \textbf{Conflicting} & \textbf{Balanced} & \textbf{Overall}\\ \hline
{\multirow{5}{*}{\textbf{GbP-Tr1}}} & G-DRO$^{\dagger}$ \cite{sagawa2020distributionally}      & 85.81$_{\pm0.16}$                   & 83.96$_{\pm0.17}$    & 84.86$_{\pm0.05}$                   & 84.93$_{\pm0.01}$    \\
{} & Vanilla   & 89.42$_{\pm0.25}$    & 77.21$_{\pm0.33}$  & 83.31$_{\pm0.05}$                   & 83.75$_{\pm0.05}$                  \\
{} & LfF \cite{nam2020learning}            & 88.73$_{\pm1.34}$                   & 77.47$_{\pm0.09}$       & 83.10$_{\pm0.64}$                   & 83.46$_{\pm0.71}$                     \\
{} & DFA \cite{lee2021learning}              & 86.12$_{\pm0.46}$                   & \textbf{77.92$_{\pm0.23}$}             & 82.02$_{\pm0.31}$                   & 82.23$_{\pm0.30}$             \\
{} & Ours                 & \textbf{90.17$_{\pm0.42}$}                   & 77.07$_{\pm1.73}$      & \textbf{83.62$_{\pm0.68}$}                   & \textbf{84.13$_{\pm0.56}$}               \\ \hline

{\multirow{5}{*}{\textbf{GbP-Tr2}}} & G-DRO$^{\dagger}$ \cite{sagawa2020distributionally}         & 83.76$_{\pm1.59}$                   & 85.14$_{\pm0.31}$      & 84.45$_{\pm0.65}$                   & 84.42$_{\pm0.61}$                             \\
{} & Vanilla           & \textbf{89.39$_{\pm0.85}$}                   & 76.13$_{\pm0.93}$      & 82.76$_{\pm0.78}$                   & 82.93$_{\pm0.78}$                       \\
{} & LfF \cite{nam2020learning}           & 87.25$_{\pm0.62}$                   & 79.07$_{\pm0.96}$      & 83.16$_{\pm0.45}$                   & 83.19$_{\pm0.44}$                           \\
{} & DFA \cite{lee2021learning}          & 80.44$_{\pm0.58}$                   & \textbf{85.51$_{\pm0.57}$}     & 82.98$_{\pm0.19}$                   & 83.09$_{\pm0.21}$                                   \\
{} & Ours           & 86.34$_{\pm0.64}$                   &  81.69$_{\pm2.67}$    & \textbf{84.02$_{\pm1.01}$}                   & \textbf{84.03$_{\pm0.97}$}       \\ \hline         

\end{tabular}}
\label{GbP_comparison}
\end{table}

\begin{figure}[b]
  \centering
    \includegraphics[width=.9\textwidth]{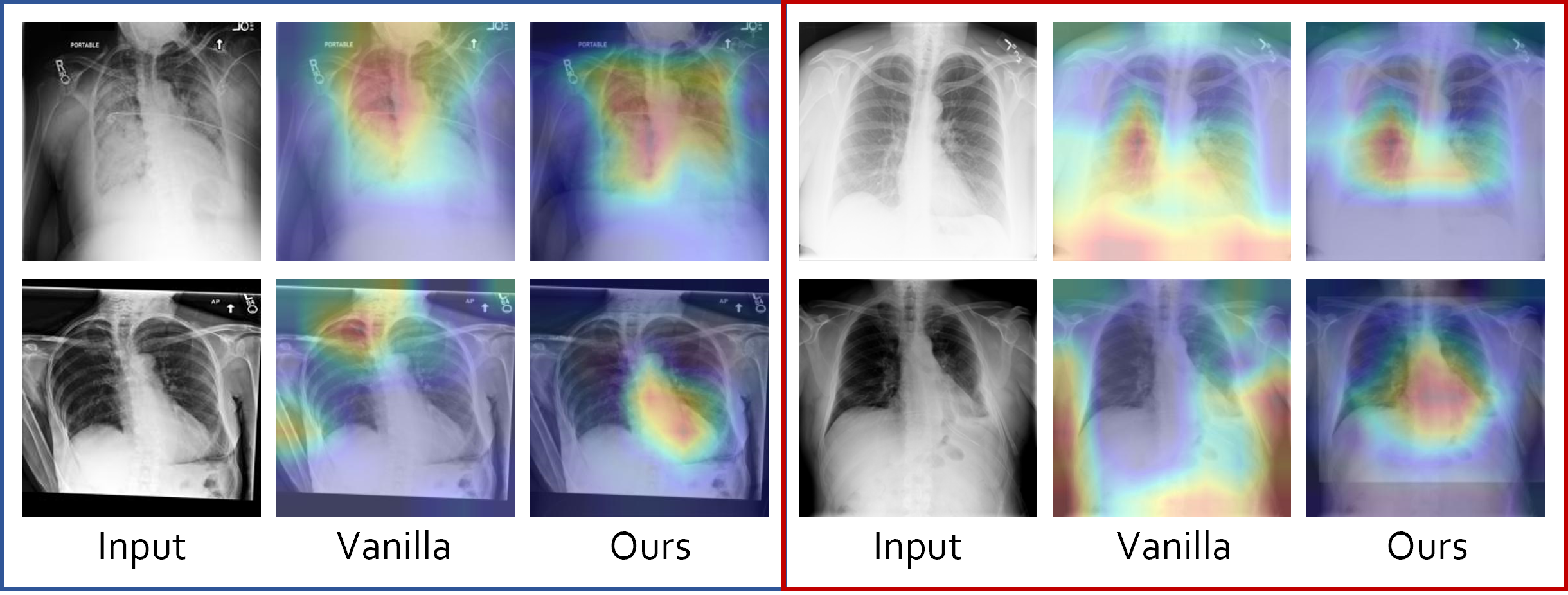}
  \caption{Class activation map \cite{zhou2016learning} generated from vanilla model and our method. Samples are from the SbP dataset (in \textcolor{Cerulean}{blue box}) and the GbP dataset (in \textcolor{BrickRed}{red box}), respectively.}
  \label{visualization}
\end{figure}

\textbf{Quantitative results on Gender-biased Pneumothorax dataset}
are reported in Table \ref{GbP_comparison}.
By the performance of the vanilla model, gender bias may not affect the performance as severely as data source bias, but it could lead to serious fairness issues.
We observed that G-DRO showed robust performance on the two different training sets.
Among approaches that do not use ground truth bias labels, our proposed method achieved consistent improvement over others with the two different training sets.
The results also showed the potential of our method in developing fair and trustworthy diagnosis models.

\textbf{Qualitative results} were visualized by class activation map \cite{zhou2016learning}, as shown in Fig. \ref{visualization}. It can be observed that vanilla model might look for evidence outside the lung regions, while our method could more correctly focus on the lung regions.

\section{Conclusion}
In this paper, we studied the causes and solutions for shortcut learning in medical image analysis, with chest X-ray as an example.
We showed that shortcut learning occurs when the bias distribution is imbalanced, and the dataset bias is preferred when it is easier to be learned than the intended features.
Based on these findings, we proposed a novel pseudo bias balanced learning algorithm to develop a debiased model without explicit labeling on the bias attribute.
We also constructed several challenging debiasing datasets from public-available data.
Extensive experiments demonstrated that our method overcame shortcut learning and achieved consistent improvements over other state-of-the-art methods under different scenarios, showing promising potential in developing robust, fair, and trustworthy diagnosis models.

\subsubsection{Acknowledgement.} This work was supported by Hong Kong Innovation and Technology Fund Project No. GHP/110/19SZ and ITS/170/20.

\bibliographystyle{paper648}
\bibliography{paper648}

\newpage

\section{Supplementary}
\subsubsection{Proof of Theorem 1} is provided following \cite{hong2021unbiased,ren2020balanced} for better reference. 
The exponential family parameterization of the multinomial distribution provides  the standard Softmax function as the \emph{canonical response function} as follows:
\begin{equation}
    \phi_{j}=\frac{{\rm exp}(\eta_{j})}{\sum_{i=1}^{k}{\rm exp}(\eta_{i})}
\end{equation}
\noindent and the \emph{canonical link function} as:
\begin{equation}
    \eta_{j} = {\rm log}(\frac{\phi_{j}}{\phi_{k}})
\label{canonical_link_function}
\end{equation}
\noindent By adding $-{\rm log}(\phi_{j}/\hat{\phi}_{j})$ to both sides of Eq. \ref{canonical_link_function}, we have:
\begin{equation}
    \eta_{j}-{\rm log}(\frac{\phi_{j}}{\hat{\phi}_{j}}) = {\rm log}(\frac{\phi_{j}}{\phi_{k}})-{\rm log}(\frac{\phi_{j}}{\hat{\phi}_{j}})={\rm log}(\frac{\hat{\phi}_{j}}{\phi_{k}}),
\end{equation}
\noindent from which we further have:
\begin{equation}
    \phi_{k}{\rm exp}(\eta_{j}-{\rm log}(\frac{\phi_{j}}{\hat{\phi}_{j}})) = \hat{\phi}_{j}
\label{phi_k_1}
\end{equation}
\begin{equation}
    \phi_{k}\sum_{i=1}^{k}{\rm exp}(\eta_{i}-{\rm log}(\frac{\phi_{i}}{\hat{\phi}_{i}})) = \sum_{i=1}^{k}\hat{\phi}_{i} = 1
\end{equation}
\begin{equation}
    \phi_{k} = 1 / \sum_{i=1}^{k}{\rm exp}(\eta_{i}-{\rm log}(\frac{\phi_{i}}{\hat{\phi}_{i}}))
\label{phi_k_3}
\end{equation}

\noindent Substitute Eq. \ref{phi_k_3} back to Eq. \ref{phi_k_1}, we could have:
\begin{equation}
    \hat{\phi}_{j} = \phi_{k}{\rm exp}(\eta_{j}-{\rm log}(\frac{\phi_{j}}{\hat{\phi}_{j}})) = \frac{{\rm exp}(\eta_{j}-{\rm log}(\frac{\phi_{j}}{\hat{\phi}_{j}}))}{\sum_{i=1}^{k}{\rm exp}(\eta_{i}-{\rm log}(\frac{\phi_{i}}{\hat{\phi}_{i}}))}
\label{321}
\end{equation}
We recall that
\begin{equation}
    \phi_{j} = p(y=j|x,b) = \frac{p(x|y=j,b)}{p(x|b)}\cdot\frac{1}{k} \text{; } \hat{\phi}_{j} = \frac{p(x|y=j,b)}{\hat{p}(x|b)}\cdot\frac{n_{j,b}}{\sum_{i=1}^{k}n_{i,b}}
\end{equation}
Hence,
\begin{equation}
    {\rm log}(\frac{\phi_{j}}{\hat{\phi}_{j}}) = {\rm log}(\frac{\sum_{i=1}^{k}n_{i,b}}{kn_{j,b}}) + {\rm log}(\frac{\hat{p}(x|b)}{p(x|b)})
\label{123}
\end{equation}
For simplicity, we let $n_{b} = \sum_{i=1}^{k}n_{j,b}$ to be the number of samples obtaining the bias label as $b$.
Finally, by substituting Eq. \ref{123} back to Eq. \ref{321}, we have
\begin{equation}
    \begin{split}
    \hat{\phi_{j}} &= \frac{{\rm exp}(\eta_{j}-{\rm log}\frac{n_{b}}{kn_{j,b}}-{\rm log}\frac{\hat{p}(x|b)}{p(x|b)})}{\sum_{i=1}^{k}{\rm exp}(\eta_{i}-{\rm log}\frac{n_{b}}{kn_{i,b}}-{\rm log}\frac{\hat{p}(x|b)}{p(x|b)})}\\
    &= \frac{\frac{n_{j,b}}{n_{b}}\cdot {\rm exp}(\eta_{j})}{\sum_{i=1}^{k}\frac{n_{i,b}}{n_{b}}\cdot {\rm exp}(\eta_{i})}
    = \frac{p(y=j|b)\cdot {\rm exp}(\eta_{j})}{\sum_{i=1}^{k}p(y=i|b)\cdot {\rm exp}({\eta_{i}})}.
\end{split}
\end{equation}

\subsubsection{Gradient of Generalized Cross Entropy Loss \cite{zhang2018generalized}:} The form of the GCE loss is as follows:
\begin{equation}
    \mathcal{L}_{\rm GCE}(f(x;\theta), y=j) = \frac{1-f_{y=j}(x;\theta)^{q}}{q},
\end{equation}
Hence, the gradient is:
\begin{equation}
    \frac{\partial \mathcal{L}_{\rm GCE}(f(x;\theta), y=j)}{\partial \theta} = -{f_{y=j}(x;\theta)}^{q-1}\frac{\partial f_{y=j}(x;\theta)}{\partial \theta}
\end{equation}
Recall that the form of conventional cross entropy loss is $\mathcal{L}_{\rm CE}(f(x;\theta), y=j) = -{\rm log}(f_{y=j}(x;\theta))$, hence
\begin{equation}
    \frac{\partial \mathcal{L}_{\rm CE}(f(x;\theta), y=j)}{\partial \theta} = -{f_{y=j}(x;\theta)}^{-1}\frac{\partial f_{y=j}(x;\theta)}{\partial\theta}
\end{equation}
Therefore, 
\begin{equation}
    \frac{\partial \mathcal{L}_{\rm GCE}(f(x;\theta), y=j)}{\partial \theta} = f_{y=j}(x;\theta)^{q}\frac{\partial \mathcal{L}_{\rm CE}(f(x;\theta), y=j)}{\partial\theta}
\end{equation}

\end{document}